\documentclass [12pt]{article}
\usepackage{graphicx}

\begin{document}

\textbf{Deuteron wave function for Reid93 potential and polarization
observables in elastic lepton-deuteron scattering}

\begin{center}
\textbf{\textit{V. I. Zhaba}}
\end{center}

\begin{center}
\textit{Uzhgorod National University, Department of Theoretical
Physics,}
\end{center}

\begin{center}
\textit{54, Voloshyna St., Uzhgorod, UA-88000, Ukraine}
\end{center}

\begin{center}
\textit{E-mail: viktorzh@meta.ua}
\end{center}

\textbf{Abstract}

Elastic lepton-deuteron scattering was considered within the limit
of zero lepton mass. The deuteron wave function in the coordinate
representation for the Reid93 potential was applied to numerical
calculations. The angular-momentum dependence of values spin
correlation coefficients $C_{xz}^{(0)}$, $C_{zz}^{(0)}$ and tensor
asymmetries $A_{xx}^{(0)}$, $A_{xz}^{(0)}$, $A_{zz}^{(0)}$ have
been evaluated in 3D format for Reid93 potential. These
polarization observables are analyzed for different energies and
scattering angles. Application of the obtained values allows
better explanation and illustration of the laws of elastic
lepton-deuteron scattering.

\textbf{Keywords}: deuteron, wave function, polarization
observables, lepton-deuteron scattering, spin correlation
coefficients, tensor asymmetries.

PACS: 13.40.Gp, 13.88.+e, 21.45.Bc, 03.65.Nk

\textbf{1. Introduction}

The interaction of three generations of leptons with deuterons has
been researched repeatedly and aroused considerable interest in
experimenters and theorists. For example, the elastic scattering
of polarized leptons (namely muons or electrons) on polarized
deuterons was investigated in \cite{Lin1970}. The differential
cross-section in terms of four form factors and the initial and
final polarizations of leptons and deuterons was described by
one-photon-exchange approximation. One of the four form factors
for describing the electromagnetic vertex of a deuteron violates
T- invariance.

Radiative Corrections to Neutrino-Deuteron Scattering are reviewed
in \cite{Kubota2006}. The problem about radiative corrections to
the SNO (i.e. Sudbury Neutrino Observatory) neutral current
induced process $\nu _{e}$+d$ \to \nu _{e}$+p+n is reexamined.
This research is related to the determination of the axial-vector
coupling constant g$_{A}$, which receives radiative corrections of
a constant term.

Shadowing in inelastic lepton-deuteron scattering is analysed in
\cite{Badelek1992} using the double interaction formalism, where
shading is associated with inclusive diffraction processes.
Different shading mechanisms are discussed in detail. The shadow
effects were very small and less than 2{\%}, according to the
precise measurements of the New Muon Collaboration.

The authors of the paper \cite{Yen1989} found that the data for
inelastic electron-deuteron scattering include some unusual
secondary peaks in the plots of deuteron inelastic structure
function \textit{vW}$_{2}$ as a function of Bjorken $x$. They
explained this effect as interference between the three-quark and
the six-quark cluster contributions to the inclusive datas.

Elastic electroweak lepton-deuteron scattering as a method to
extract information about vector and axial-vector isoscalar
currents, and thus the strange and heavier quark content in the
deuteron was considered in paper \cite{Pollock1990}.

In paper \cite{Yen1990} was appreciated the contribution of
quasifree nucleon knockout and of inelastic lepton-nucleon
scattering in inclusive electron-deuteron reactions at large
momentum transfer. The degree of quantitative agreement with
deuteron wave functions for the Reid soft-core and Bonn potentials
was investigated. In the range of data's is contained strong
sensitivity to the tensor correlations, which are visibly
different in these two deuteron models. Here the deuteron wave
function for Reid soft-core potential describes the data better
than for the Bonn potential. A good description of all data with
both these nucleon-nucleon interactions was obtained by the
inclusion of a six-quark cluster component, whose relative
contribution is based on the overlapping criterion.

The energy and angular distributions of slow nucleons in
semi-inclusive inelastic lepton scattering off the deuteron are
investigated in paper \cite{Simula1998}. The spectator scaling
property for semi-inclusive cross-section within the spectator
mechanism is an interesting instrument to obtain model-independent
information on the neutron structure function in the
deep-inelastic regime and in the regions of nucleon-resonance
productions.

For the x-rescaling model \cite{Kaptari1992} in deep inelastic
lepton scattering on bound nucleons has been applied the operator
product expansion method within the effective meson-nucleon
theory. Demonstrated that with the such contributions as Fermi
motion and mesonic corrections, the x-rescaling idea is exactly
reproduced.

An analysis of polarization observables \cite{Kaptari1995} in two
such processes, as the deep-inelastic scattering of polarized
leptons off polarized deuterons and the polarized deuteron
break-up, indicates a relation between the deep-inelastic
structure functions $b_{1,2}(x)$ and the tensor analyzing power
$T_{20}$.

In paper \cite{Bondarenko2002} it is shown that the Bethe-Salpeter
approach with the use of the separable interaction for the
deuteron allows a covariant description of different
electromagnetic reactions such as the lepton-deuteron scattering,
deuteron electro-disintegration, deep inelastic scattering of
leptons on light nuclei.

Model independent expressions for polarization observables in
elastic lepton scattering on the proton are obtained in
\cite{Gakh2015}, taking into account the lepton mass and the
two-photon exchange contribution. General information about the
influence of the two-photon-exchange expressions on the
differential cross-section and on polarization observables are
given. Polarization effects have also been investigated for the
case of a longitudinally polarized lepton beam and polarized
nucleon in the final state.

The authors of paper \cite{Eskin2017} evaluated the contribution
to the polarizability of the nucleus to hyperfine structure of
muonic hydrogen within the framework of unitary isobar model and
on the basis of experimental data's on the structure functions of
deep inelastic lepton-proton and lepton-deuteron scatterings. The
calculations of virtual absorption cross-sections of transversely
and longitudinally polarized photons by nucleons in the resonance
area were made in the program MAID.

The differential cross-section including quantum electrodynamics
(QED) radiative corrections to the leptonic part of the
interaction, in the case of a coincidence experimental setup is
derived in paper \cite{Gakh2018}.

In this paper we use the analytic forms of DWF in coordinate space for
theoretical calculations of three main sets of polarization observables in
elastic lepton-deuteron scattering taking into account the limits of zero
lepton mass. Nucleon-nucleon realistic phenomenological potential of
Nijmegen group Reid93 are used for numerical calculations.

\textbf{2. Deuteron wave function}

The deuteron wave function (DWF) in coordinate space can be
presented as a table: through respective two arrays of values of
radial wave functions $u(r)$ and $w(r)$. It is heavy to apply such
arrays for practical numerical calculations. That is why use
simple formulas for analytical forms of DWF representation
\cite{MPLA2016}.

The known numerical values of DWFs in coordinate space can be approximated
using expansions in the convenient analytical form:

1) for Paris potential \cite{Lacombe1981}

\begin{equation}
\label{eq1}
\left\{ {\begin{array}{l}
 u\left( r \right) = \sum\limits_{j = 1}^N {C_j \exp \left( { - m_j r}
\right),} \\
 w\left( r \right) = \sum\limits_{j = 1}^N {D_j \exp \left( { - m_j r}
\right)\left[ {1 + \frac{3}{m_j r} + \frac{3}{\left( {m_j r} \right)^2}}
\right],} \\
 \end{array}} \right.
\end{equation}

where $m_j = \beta + (j - 1)m_0 $; $\beta = \sqrt {ME_d } $, $m_{0}$=0.9
fm$^{ - 1}$; $M$ -- nucleon mass; $E_{d}$ -- binding energy of the deuteron;

2) for Moscow potential \cite{Krasnopolsky1985} and the dressed
dibaryon model (DDM) \cite{Platonova2010}:

\begin{equation}
\label{eq2}
\left\{ {\begin{array}{l}
 u(r) = r\sum\limits_{i = 1}^N {A_i \exp ( - a_i r^2),} \\
 w(r) = r^3\sum\limits_{i = 1}^N {B_i \exp ( - b_i r^2).} \\
 \end{array}} \right.
\end{equation}

3) for Nijmegen group potentials (NijmI, NijmII, Nijm93, Reid93)
and Argonne v18 potential \cite{NPAE2016, MPLA2016}:

\begin{equation}
\label{eq3}
\left\{ {\begin{array}{l}
 u(r) = r^{3 / 2}\sum\limits_{i = 1}^N {A_i \exp ( - a_i r^3),} \\
 w(r) = r\sum\limits_{i = 1}^N {B_i \exp ( - b_i r^3).} \\
 \end{array}} \right.
\end{equation}

The analytical form (\ref{eq1}) was also applied to OBEPC
\cite{Machleidt1989} and CD-Bonn \cite{Machleidt2001} potentials
and fss2 \cite{Fujiwara2001} and MT \cite{Krutov2007} models.
Advantage of analytical form (\ref{eq3}) consists in that
calculated DWFs do not contain superfluous knots.

3. \textbf{Elastic lepton-deuteron scattering}

The expressions for the unpolarized differential cross-section and
polarization observables were studied in the application of the
lepton mass in elastic lepton-deuteron scattering by
\cite{Gakh2014}. The asymmetries conditioned to the tensor
polarization of the deuteron target and the spin correlation
coefficients conditioned to the lepton beam polarization and
vector polarization of the deuteron target has been studied.

The lepton-deuteron reaction is written down in a form

\begin{equation}
\label{eq4}
l(k_1 ) + d(p_1 ) \to l(k_2 ) + d(p_2 ),
\quad
l = e,\mu ,\tau .
\end{equation}

Such reaction in the laboratory system is characterized by 4-
moments for a deuteron (lepton) in initial and final states of
$p_{1}$ and $p_{2}$ ($k_{1}$ and $k_{2})$ with the corresponding
components

\[
p_1 = (M_D ,0), \quad p_2 = (E_2 ,\vec {p}_2 ), \quad k_1 =
(\varepsilon _1 ,\vec {k}_1 ), \quad k_2 = (\varepsilon _2 ,\vec
{k}_2 ),
\]

where $M_D $ is the deuteron mass.

Comparison of formulas for cross-sections $\sigma _{MOTT} $ in
\cite{Gakh2014}

\[
\sigma _0 (m = 0) = \sigma _{MOTT} = \frac{\alpha ^2\cos ^2\left(
{\frac{\theta }{2}} \right)}{4\varepsilon _1^2 \sin ^4\left( {\frac{\theta
}{2}} \right)}\left( {1 + \frac{2\varepsilon _1 }{M_D }\sin ^2\left(
{\frac{\theta }{2}} \right)} \right)^{ - 1}
\]

and in \cite{Garcon1994}

\[
\left( {\frac{d\sigma }{d\Omega }} \right)_{MOTT} = \frac{1}{f}\frac{\alpha
^2\cos ^2\left( {\frac{\theta _e }{2}} \right)}{4E_e^2 \sin ^4\left(
{\frac{\theta _e }{2}} \right)},
\]

allows you to determine the value $\varepsilon _1 $.

According to \cite{Garcon1994}, the factor in the cross-section is
presented as

\[
f = 1 + \frac{2E_e }{M_D }\sin ^2\left( {\frac{\theta _e }{2}} \right).
\]

That's why in \cite{Gakh2014} $\varepsilon _1 = E_e $. The
relation between momentum and energy is given by the expression

\[
p^2 = 4\frac{E_e^2 }{f}\sin ^2\left( {\frac{\theta _e }{2}} \right).
\]

Thus, the following relations can be written for momentum and energy

\[
p = \sqrt {\frac{4\varepsilon _1^2 \sin ^2\left( {\frac{\theta }{2}}
\right)}{1 + \frac{2\varepsilon _1 }{M_D }\sin ^2\left( {\frac{\theta }{2}}
\right)}} ;
\]

\[
\varepsilon _1 = \frac{p^2 + \csc \left[ {\frac{\theta }{2}} \right]^2\sqrt
{p^4\sin ^4\left[ {\frac{\theta }{2}} \right] + 4p^2\sin ^2\left[
{\frac{\theta }{2}} \right]M_D^2 } }{4M_D }.
\]

That is, the value $\varepsilon _1 $ in the dimensions [fm$^{-1}$]
or [GeV/c].

If known DWFs are in the coordinate representation, then numerically we can
calculate the polarization observed in the elastic lepton-deuteron
scattering. What types of polarization observables can be calculated? Let's
consider three basic sets of polarization characteristics in elastic
lepton-deuteron scattering.

We first consider the vector-polarized deuteron target. The differential
cross-section for the reaction (\ref{eq4}) describes the scattering of polarized
lepton beam on the vector-polarized deuteron target

\begin{equation}
\label{eq5}
\frac{d\sigma (s,s_l )}{d\Omega } = \frac{d\sigma _{un} }{d\Omega }\left( {1
+ C_{xx} \xi _x \xi _{lx} + C_{yy} \xi _y \xi _{ly} + C_{zz} \xi _z \xi
_{lz} + C_{xz} \xi _x \xi _{lz} + C_{zx} \xi _z \xi _{lx} } \right),
\end{equation}

where $\vec {\xi }_l $ and $\vec {\xi }$ is the unit polarization vectors in
the rest frame of the lepton beam and deuteron target respectively. The
expression (\ref{eq5}) describes only to the spin-dependent part of the
cross-section which is determined by the spin correlation coefficients.

Here coefficients characterize the scattering of the longitudinally
polarized lepton beam, and coefficients $C_{xx}^{(0)} $, $C_{yy}^{(0)} $,
$C_{zx}^{(0)} $ correspond to the transverse components of the spin vector
$\vec {\xi }_l $.

The spin correlation coefficients $C_{ij}$ in the limit of zero
lepton mass have the such form in terms of the deuteron
electromagnetic form factors \cite{Gakh2014}:

\begin{equation}
\label{eq6}
\bar {D}C_{xz}^{(0)} = \frac{1}{2}\frac{\tau }{\varepsilon _1 }tg\left(
{\frac{\theta }{2}} \right)\left[ {(\varepsilon _1 + \varepsilon _2 )G_M -
4(M + \varepsilon _1 )\left( {G_C + \frac{\tau }{3}G_Q } \right)} \right]G_M
;
\end{equation}

\begin{equation}
\label{eq7}
\bar {D}C_{zz}^{(0)} = - 2\tau G_M \frac{M}{\varepsilon _1 }\left[ {G_C +
\frac{\tau }{3}G_Q + \frac{\varepsilon _2 }{2M^2}(M + \varepsilon _1 )\left(
{1 + \frac{\varepsilon _1 }{M}\sin ^2\left( {\frac{\theta }{2}} \right)}
\right)tg^2\left( {\frac{\theta }{2}} \right)G_M } \right];
\end{equation}

\begin{equation}
\label{eq8}
C_{xx}^{(0)} = C_{yy}^{(0)} = C_{zx}^{(0)} = 0;
\end{equation}

where $\bar {D} = A(p) + B(p)tg^2\left( {\frac{\theta }{2}}
\right)$ is factor, which is determined by the structure functions
$A$ and $B$; the charge $G_{C}(p)$, quadrupole $G_{Q}(p)$
and magnetic $G_{M}(p)$ form factors
contain information about the electromagnetic properties of the deuteron;
$\tau = \frac{p^2}{4M_D^2 }$.

The connection between $\varepsilon _1 $ and $\varepsilon _2 $ is written as

\[
\varepsilon _2 = \frac{\varepsilon _1 }{1 + \frac{2\varepsilon _1 }{M_D
}\sin ^2\left( {\frac{\theta }{2}} \right)}.
\]

For the description of the elastic lepton-deuteron scattering it is possible
to use other system of coordinates with such axes: the Z and Y axes is
directed along the virtual photon momentum and along the vector $\vec {k}_1
\times \vec {k}_2 $ respectively; axis X is chosen for formation of the
left-handed coordinate system. The reaction scattering plane is
characterized by a angle $\psi $ between the direction of the lepton beam
and the virtual photon momentum

\[
\cos \psi = \frac{M_D + \varepsilon _1 }{\left| {\vec {k}_1 } \right|}\sqrt
{\frac{\tau }{1 + \tau }} ;
\quad
\sin \psi = - \frac{1}{\left| {\vec {k}_1 } \right|\sqrt {1 + \tau } }\sqrt
{\varepsilon _1 \varepsilon _2 - \tau M_D^2 - m^2(1 + \tau )} .
\]

Formulas for spin correlation coefficients in the new coordinate system can
be obtained in terms for coefficients (\ref{eq6})-(\ref{eq7})

\[
C_{Zz} = \cos \psi C_{zz} + \sin \psi C_{xz} ;
\quad
C_{Xz} = - \sin \psi C_{zz} + \cos \psi C_{xz} ;
\]

\[
C_{Zx} = \cos \psi C_{zx} + \sin \psi C_{xx} ;
\quad
C_{Xx} = - \sin \psi C_{zx} + \cos \psi C_{xx} .
\]

In such representation of coordinates are present only two nonzero spin
correlation coefficients in the limit of zero lepton mass

\begin{equation}
\label{eq9}
\bar {D}C_{Zz}^{(0)} = - \tau \sqrt {(1 + \tau )\left( {1 + \tau \sin
^2\left( {\frac{\theta }{2}} \right)} \right)} tg\left( {\frac{\theta }{2}}
\right)\sec \left( {\frac{\theta }{2}} \right)G_M^2 ;
\end{equation}

\begin{equation}
\label{eq10}
\bar {D}C_{Xz}^{(0)} = - 2\sqrt {\tau (1 + \tau )} tg\left( {\frac{\theta
}{2}} \right)G_M \left( {G_C + \frac{\tau }{3}G_Q } \right).
\end{equation}

It should be noted that the values $C_{Zz}^{(0)} $, $C_{Xz}^{(0)}
$ are the same quantities $A_B^L $, $A_B^T $ as in
\cite{Gakh2012}. The spin correlation coefficients
(\ref{eq9})-(\ref{eq10}) correspond to the longitudinal
polarization of the lepton beam.

We will also consider the tensor-polarized deuteron target. The differential
cross-section for the reaction (\ref{eq4}) describes the scattering of unpolarized
lepton beam on such tensor-polarized deuteron target

\[
\frac{d\sigma (s,s_l )}{d\Omega } = \frac{d\sigma _{un} }{d\Omega }\left[ {1
+ A_{xx} (Q_{xx} - Q_{yy} ) + A_{xz} Q_{xz} + A_{zz} Q_{zz} } \right],
\]

where it is used conditions that $Q_{xx} + Q_{yy} + Q_{xz} = 0$.
Here $A_{ij}$ - the asymmetries induced by the tensor polarization
of the deuteron target. The asymmetries in the limit of zero
lepton mass is written in terms of the deuteron electromagnetic
form factors \cite{Gakh2014}:

\begin{equation}
\label{eq11}
\begin{array}{l}
 \bar {D}A_{xx}^{(0)} = \frac{\tau }{2}\left\{ {\left( {1 + \tau
\frac{M^2}{\varepsilon _1^2 }} \right)} \right.G_M^2 + \frac{4}{1 + \tau
}G_Q \left[ {\tau \left( {1 + \frac{M}{\varepsilon _1 }} \right)\left( {1 -
\tau \frac{M}{\varepsilon _1 }} \right)G_M + } \right. \\
 \left. {\left. { + \left( {1 - \tau \frac{M^2}{\varepsilon _1^2 } - 2\tau
\frac{M}{\varepsilon _1 }} \right)\left( {G_C + \frac{\tau }{3}G_Q }
\right)} \right]} \right\} ; \\
 \end{array}
\end{equation}

\begin{equation}
\label{eq12}
\begin{array}{l}
 \bar {D}A_{xz}^{(0)} = - \frac{\varepsilon _2 }{M}\frac{\tau }{1 + \tau
}\sin \theta \left\{ {4\left( {1 + \frac{M}{\varepsilon _1 }} \right)G_Q
\left( {G_C + \frac{\tau }{3}G_Q } \right) + (1 + \tau )\left( {1 +
\frac{M}{\varepsilon _1 }} \right)tg^2\left( {\frac{\theta }{2}}
\right)G_M^2 } \right. + \\
 \left. { + 2\left( {1 - \tau \frac{M}{\varepsilon _1 }} \right)\left[ { - 1
- \tau + 2\sin ^2\left( {\frac{\theta }{2}} \right)\left( {1 +
\frac{\varepsilon _1 }{M} + \frac{\varepsilon _1^2 }{M^2} - \tau
\frac{\varepsilon _1 }{M}} \right)} \right]\left( {1 + tg^2\left(
{\frac{\theta }{2}} \right)} \right)G_M G_Q } \right\} ; \\
 \end{array}
\end{equation}

\begin{equation}
\label{eq13}
\begin{array}{l}
 \bar {D}A_{zz}^{(0)} = - \frac{\tau }{2}\left\{ {\left[ {6\frac{\tau }{1 +
\tau }\frac{\varepsilon _1 + \varepsilon _2 }{\varepsilon _1 }\left( {1 +
\frac{M}{\varepsilon _1 }} \right)G_Q - G_M } \right]} \right.G_M + \\
 + tg^2\left( {\frac{\theta }{2}} \right)\left[ {1 - 2\tau - 6\tau
\frac{M}{\varepsilon _1 }\left( {1 + \frac{M}{2\varepsilon _1 }} \right)}
\right]\left. {\left[ {G_M^2 + \frac{4}{1 + \tau }\cot ^2\left(
{\frac{\theta }{2}} \right)G_Q \left( {G_C + \frac{\tau }{3}G_Q } \right)}
\right]} \right\} . \\
 \end{array}
\end{equation}

If the $Z$ axis is directed along the virtual photon momentum,
then the asymmetries due to the tensor polarization of the
deuteron target have such a form \cite{Gakh2014}

\[
A_\alpha = T_{\alpha \beta } (\psi )A^\beta .
\]

Here \textit{$\alpha $}=\textit{ZZ}, \textit{XX}, \textit{XZ} and \textit{$\beta $}=\textit{zz}, \textit{xx}, \textit{xz} - the indices of the rotation matrix. The rotation
matrix write as

\begin{equation}
\label{eq14}
T(\psi ) = \left( {{\begin{array}{*{20}c}
 {\frac{1}{4}\left( {1 + 3\cos (2\psi )} \right)} \hfill &
{\frac{3}{4}\left( {1 - \cos (2\psi )} \right)} \hfill & {\frac{3}{4}\sin
(2\psi )} \hfill \\
 {\frac{1}{4}\left( {1 - \cos (2\psi )} \right)} \hfill & {\frac{1}{4}\left(
{3 + \cos (2\psi )} \right)} \hfill & { - \frac{1}{4}\sin (2\psi )} \hfill
\\
 { - \sin (2\psi )} \hfill & {\sin (2\psi )} \hfill & {\cos (2\psi )} \hfill
\\
\end{array} }} \right).
\end{equation}

The transformation for the tensor of the quadrupole polarization describes
the tensor polarization of the deuteron target

\[
Q_{ZZ} = \frac{1}{4}\left( {1 + 3\cos (2\psi )} \right)Q_{zz} +
\frac{1}{4}\left( {1 - \cos (2\psi )} \right)(Q_{xx} - Q_{yy} ) + \sin
(2\psi )Q_{xz} ;
\]

\[
Q_{XX} - Q_{YY} = \frac{3}{4}\left( {1 - \cos (2\psi )} \right)Q_{zz} +
\frac{1}{4}\left( {3 + \cos (2\psi )} \right)(Q_{xx} - Q_{yy} ) + \sin
(2\psi )Q_{xz} ;
\]

\[
Q_{XZ} = - \frac{3}{4}\sin (2\psi )Q_{zz} + \frac{1}{4}\sin (2\psi )(Q_{xx}
- Q_{yy} ) + \cos (2\psi )Q_{xz} .
\]

Consequently, within the limits of zero lepton mass we considered three main
sets of polarization characteristics in elastic lepton-deuteron scattering:

1) the spin correlation coefficients (\ref{eq6})-(\ref{eq7});

2) two nonzero spin correlation coefficients (\ref{eq9})-(\ref{eq10});

3) the tensor asymmetries (\ref{eq11})-(\ref{eq13}).

All of the aforementioned formulas for an unpolarized differential
cross-section and polarization observables with consideration the
lepton mass for elastic lepton-deuteron scattering are cited from
paper \cite{Gakh2014}. There are described expressions for
asymmetries owing to the tensor polarization of the deuteron
target and the spin correlation coefficients owing to the lepton
beam polarization and vector polarization of the deuteron target.
Explicit, ostensive and understandable formulas for these
quantities have been demonstrated in two coordinate systems: in
the first one system the z- axis is directed along the lepton beam
momentum and in the second one system the z- axis is directed
along the virtual photon momentum (or the transferred momentum).
Such coordinate systems are necessary and relevant for
experimental research.

\textbf{4. Calculations and conclusions}

We will analyze numerical results for two-dimensional plots with function of
the muon beam energy and the muon scattering angle, that determine the
kinematics for a binary process.

The values of spin correlation coefficients $C_{xz}^{(0)}
(\varepsilon _1 ,\theta )$, $C_{zz}^{(0)} (\varepsilon _1 ,\theta
)$ and tensor asymmetries $A_{xx}^{(0)} (\varepsilon _1 ,\theta
)$, $A_{xz}^{(0)} (\varepsilon _1 ,\theta )$, $A_{zz}^{(0)}
(\varepsilon _1 ,\theta )$ (the formulas (\ref{eq6}), (\ref{eq7})
and (\ref{eq11})-(\ref{eq13}) respectively) for elastic
lepton-deuteron scattering in the limit of zero lepton mass are
calculated. The results of numerical calculations are presented in
3D format in Figs. 1-5. Nucleon-nucleon realistic phenomenological
potential Reid93 was used for calculations. A complete set of
coefficients for DWF (\ref{eq3}) for Reid93 potential is given in
paper \cite{MPLA2016}.

\pdfximage width 90mm {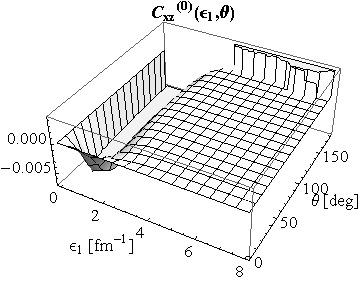}\pdfrefximage\pdflastximage

Fig. 1. Spin correlation coefficient $C_{xz}^{(0)} (\varepsilon _1
,\theta )$

\pdfximage width 90mm {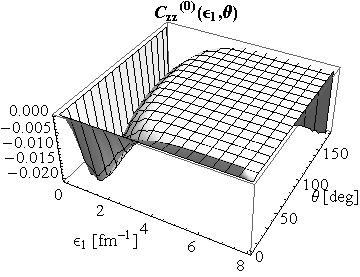}\pdfrefximage\pdflastximage

Fig. 2. Spin correlation coefficient $C_{zz}^{(0)} (\varepsilon _1 ,\theta
)$

\pdfximage width 90mm {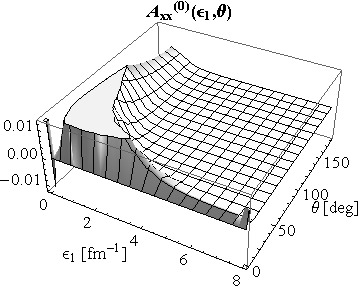}\pdfrefximage\pdflastximage

Fig. 3. Tensor asymmetry $A_{xx}^{(0)} (\varepsilon _1 ,\theta )$

\pdfximage width 90mm {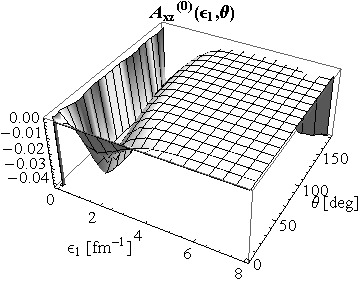}\pdfrefximage\pdflastximage

Fig. 4. Tensor asymmetry $A_{xz}^{(0)} (\varepsilon _1 ,\theta )$

\pdfximage width 90mm {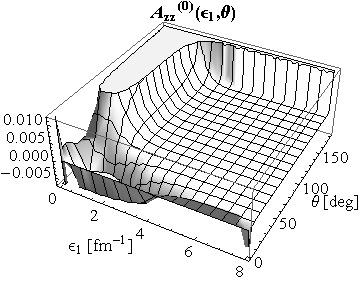}\pdfrefximage\pdflastximage

Fig. 5. Tensor asymmetry $A_{zz}^{(0)} (\varepsilon _1 ,\theta )$

Figs. 1 and 2 illustrates the polarization observables $C_{xz}^{(0)}
(\varepsilon _1 ,\theta )$, $C_{zz}^{(0)} (\varepsilon _1 ,\theta )$, which
are induced by a polarized lepton beam on a vector-polarized deuteron
target. These spin correlation coefficients vanish at zero angle scattering
and at small energies. They become negative and forms a kind of ``tray'' as
values of the angle and energy increase.

The values $C_{Zz}^{(0)}$, $C_{Xz}^{(0)}$ for Reid93 potential
were calculated in the paper \cite{ZhabaWSN114}, where they are
labeled as $A_B^L$, $A_B^T$.

The tensor asymmetries $A_{xx}^{(0)} (\varepsilon _1 ,\theta )$,
$A_{xz}^{(0)} (\varepsilon _1 ,\theta )$, $A_{zz}^{(0)}
(\varepsilon _1 ,\theta )$ induced by a unpolarized lepton beam on
tensor-polarized deuteron target are illustrated in Fig. 3-5
respectively. For tensor asymmetries the characteristic plane at
energy increase above 4 fm$^{-1}$. As seen in Figs. 3 and 5 for
tensor asymmetries $A_{xx}^{(0)}$ and $A_{zz}^{(0)}$ there is a
hump (peak) near 2 fm$^{-1}$ in the range of angles 0-180 degrees.
For the tensor asymmetry $A_{xz}^{(0)}$ (see. Fig. 4), on the
contrary, there is a pit.

Perspectives are the following numeric calculations of the unpolarized
cross-section, spin correlation coefficients and tensor asymmetries taking
into account the lepton mass. Then it is possible to compare them with the
results within the limit of zero lepton mass which are obtained in this
paper.

It is interesting to use different parametrizations of the
deuteron form factors, the form of which in fact determines the
behavior of the spin correlation coefficients and tensor
asymmetries. As indicated in the paper \cite{Gakh2014}, the chosen
model for deuteron form factors should accurately reproduce the
existing experimental data, and the effect of the finite lepton
mass is sizable at low incident energies and large scattering
angles.

Considering and neglecting the lepton mass will better describe
proton-antiproton annihilation into massive leptons and
polarization phenomena \cite{Dbeyssi2012}.

\end{document}